\documentclass[10pt, paper=a4, UKenglish]{article}
\usepackage{graphicx}
\def\Title#1{\begin{center} {\Large #1 } \end{center}}
\def\Author#1{\begin{center}{ \sc #1} \end{center}}
\def\Address#1{\begin{center}{ \it #1} \end{center}}

\newcommand\pubblock{\rightline{\begin{tabular}{l} Proceedings of CTD 2020\\ \pubnumber\\
         \pubdate  \end{tabular}}}

\newenvironment{Abstract}{\begin{quotation} \begin{center} 
             \large ABSTRACT \end{center}\bigskip 
      \begin{center}\begin{large}}{\end{large}\end{center} \end{quotation}}

\newenvironment{Presented}{\begin{quotation} \begin{center} 
             PRESENTED AT\end{center}\bigskip 
      \begin{center}\begin{large}}{\end{large}\end{center} \end{quotation}}

\def\Acknowledgements{\bigskip  \bigskip \begin{center} \begin{large}
      \bf ACKNOWLEDGMENTS \end{large}\end{center}}

\def\beq{\begin{equation}}
\def\eeq#1{\label{#1}\end{equation}}
\def\eeqn{\end{equation}}

\def\beqa{\begin{eqnarray}}
\def\eeqa#1{\label{#1}\end{eqnarray}}
\def\eeqan{\end{eqnarray}}

\let\bar=\overbar

\def\Dslash{\not{\hbox{\kern-4pt $D$}}}
\def\dslash{\not{\hbox{\kern-2pt $\del$}}}

\def\msb{{\bar{\ssstyle M \kern -1pt S}}}

\usepackage{color}
\usepackage{lineno}
\usepackage{subfig}
\usepackage{hyperref}

\newcommand\pubnumber{PROC-CTD2020-61}

\newcommand\pubdate{\today}

\def\affiliation{
Oak Ridge National Laboratory \\
Oak Ridge, Tennessee}

\newcommand{\conference}{Connecting the Dots Workshop (CTD 2020)\\
April 20-30, 2020}

\usepackage{fancyhdr}
\definecolor{mygrey}{RGB}{105,105,105}
\begin{document}


\large
\begin{titlepage}
\pubblock

\vfill
\Title{Requirements, Status, and Plans for Track Reconstruction at the sPHENIX Experiment}
\vfill

\Author{Joseph D. Osborn for the sPHENIX Collaboration}
\Address{\affiliation}
\vfill

\begin{Abstract}
sPHENIX is a new experiment that is being constructed at the Relativistic Heavy Ion Collider at Brookhaven National Laboratory. The primary physics goals of sPHENIX are to measure jets, their substructure, and the upsilon resonances in $p$$+$$p$, $p$+Au and Au+Au collisions. To realize these goals, a tracking system composed of a time projection chamber and several silicon detectors will be used to identify tracks that correspond to jets and upsilon decays. However, the sPHENIX experiment will collect approximately 200 PB of data utilizing a finite-sized computing center; thus, performing track reconstruction in a timely manner is a challenge due to the large occupancy of heavy-ion collisions. The sPHENIX experiment, its track reconstruction, and the need for implementing faster track -fitting algorithms, such as that provided by the A Common Tracking Software package, into the sPHENIX software stack are discussed.
\end{Abstract}

\vfill

\begin{Presented}
\conference
\end{Presented}
\vfill
\end{titlepage}
\def\thefootnote{\fnsymbol{footnote}}
\setcounter{footnote}{0}
%

\normalsize 

\section*{Notice of Copyright}
This manuscript has been authored by UT-Battelle LLC, under contract DE-AC05-00OR22725 with the US Department of Energy (DOE). The US government retains and the publisher, by accepting the article for publication, acknowledges that the US government retains a nonexclusive, paid-up, irrevocable, worldwide license to publish or reproduce the published form of this manuscript, or allow others to do so, for US government purposes. DOE will provide public access to these results of federally sponsored research in accordance with the DOE Public Access Plan (http://energy.gov/downloads/doe-public-access-plan).

\section{Introduction}
\label{intro}

The sPHENIX experiment is a next-generation jet and heavy flavor detector being constructed for operation at the Relativstic Heavy Ion Collider (RHIC) at Brookhaven National Laboratory (BNL)~\cite{sphenix}. The primary physics goal for sPHENIX is to study strong force interactions by probing the inner workings of the quark-gluon-plasma (QGP) created in heavy nucleus-nucleus collisions, as outlined in the 2015 Nuclear Physics Long-Range Plan~\cite{LRP}. sPHENIX will also probe the structure of protons and nuclei in proton-proton and proton-nucleus collisions~\cite{sphenixcoldqcd}. To achieve these goals, the detector is designed as a precision jet and heavy-flavor spectrometer. Jets, and their structure, can resolve strong force interactions at different scales when parton flavor is selected due to the difference in mass between heavy and light quarks. Similarly, the measurement of $\Upsilon(1S)$ and its first two excited states allow different screening temperatures of the QGP to be accessed. To achieve these physics goals, precise tracking capabilities are required.

However, accomplishing the precision tracking that is required for these measurements in the environment that RHIC will provide to sPHENIX will be a significant challenge. The accelerator will deliver Au+Au nucleus collisions at $\sqrt{s_{_{NN}}}=200$ GeV at a rate of up to $\sim$200 kHz, which means that sPHENIX will experience between three to eight pileup events per bunch crossing. Central heavy ion collisions can produce approximately 1,000 particles per event; therefore, the occupancy in the detector will be extremely high. Over a 3 year running period, while sPHENIX collects data at a rate of 15 kHz, these conditions will lead to an accumulation of nearly 250 PB of data. The data processing will also be performed on a finite-sized computing center at BNL. Therefore, sPHENIX requires high-speed, efficient, and precise tracking in an environment in which $\mathcal{O}$(100,000) hits are expected within the tracking detector volume. To process and analyze the data in a timely fashion given these constraints, the track reconstruction software must be able to track an entire event within a 5 second HS06 CPU core  budget. To reach this goal, A Common Tracking Software (ACTS)~\cite{acts, actstext} is being implemented into the sPHENIX software stack. 

\section{sPHENIX Detector and Physics Requirements}
\label{geometry}

\begin{figure}[!htb]
	\centering
	\includegraphics[width=0.6\linewidth]{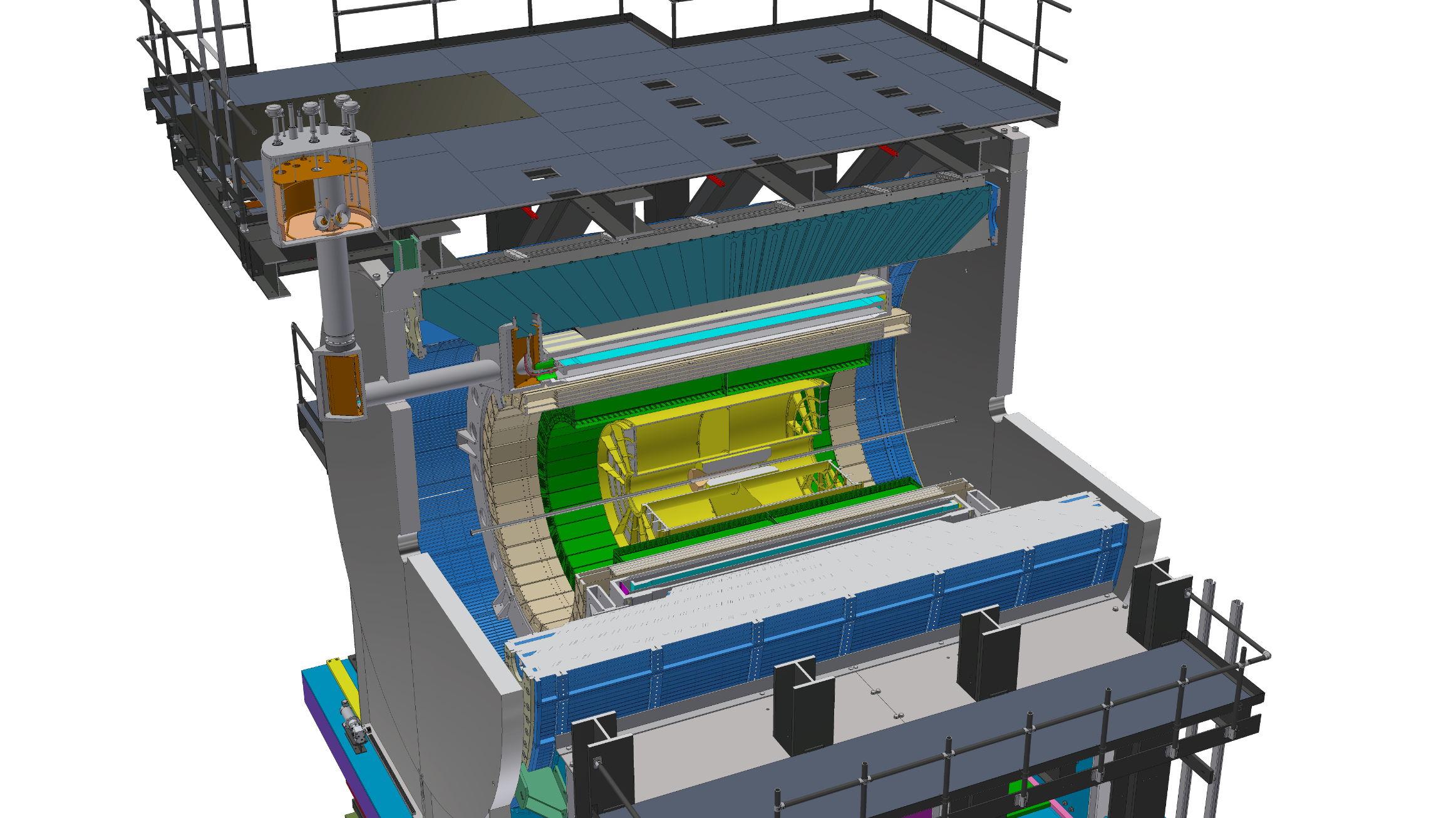}
	\caption{An engineering diagram of the sPHENIX detector design. The MVTX and INTT are two subdetectors that are composed of silicon staves, shown in orange and grey, respectively. The TPC is a continuous readout GEM-based detector, and the TPC cage is shown in yellow.}
	\label{fig:detector}
\end{figure}

The sPHENIX spectrometer is a midrapidity barrel detector with full azimuthal and pseudorapidity $|\eta|<1.1$ acceptance, and includes tracking and electromagnetic and hadronic calorimeters. An engineering drawing of the detector is shown in Fig.~\ref{fig:detector}. The primary tracking detectors are a monolothic active pixel sensor (MAPS) based vertex detector (MVTX), the Intermediate Tracker (INTT), and a time projection chamber (TPC). The MVTX has three layers of silicon staves that cover a radial distance of approximately $2<r<4$ cm from the beam pipe. The INTT has two layers of silicon strips and covers approximately $7<r<10$ cm. The TPC is the primary tracking detector within sPHENIX and is a compact, continuous readout gas electron multiplier (GEM)-based TPC. These detectors are described in greater detail in the sPHENIX Technical Design Report~\cite{sphenixtdr}.

\begin{figure}[!tbh]
	\centering
	\includegraphics[width=0.5\linewidth]{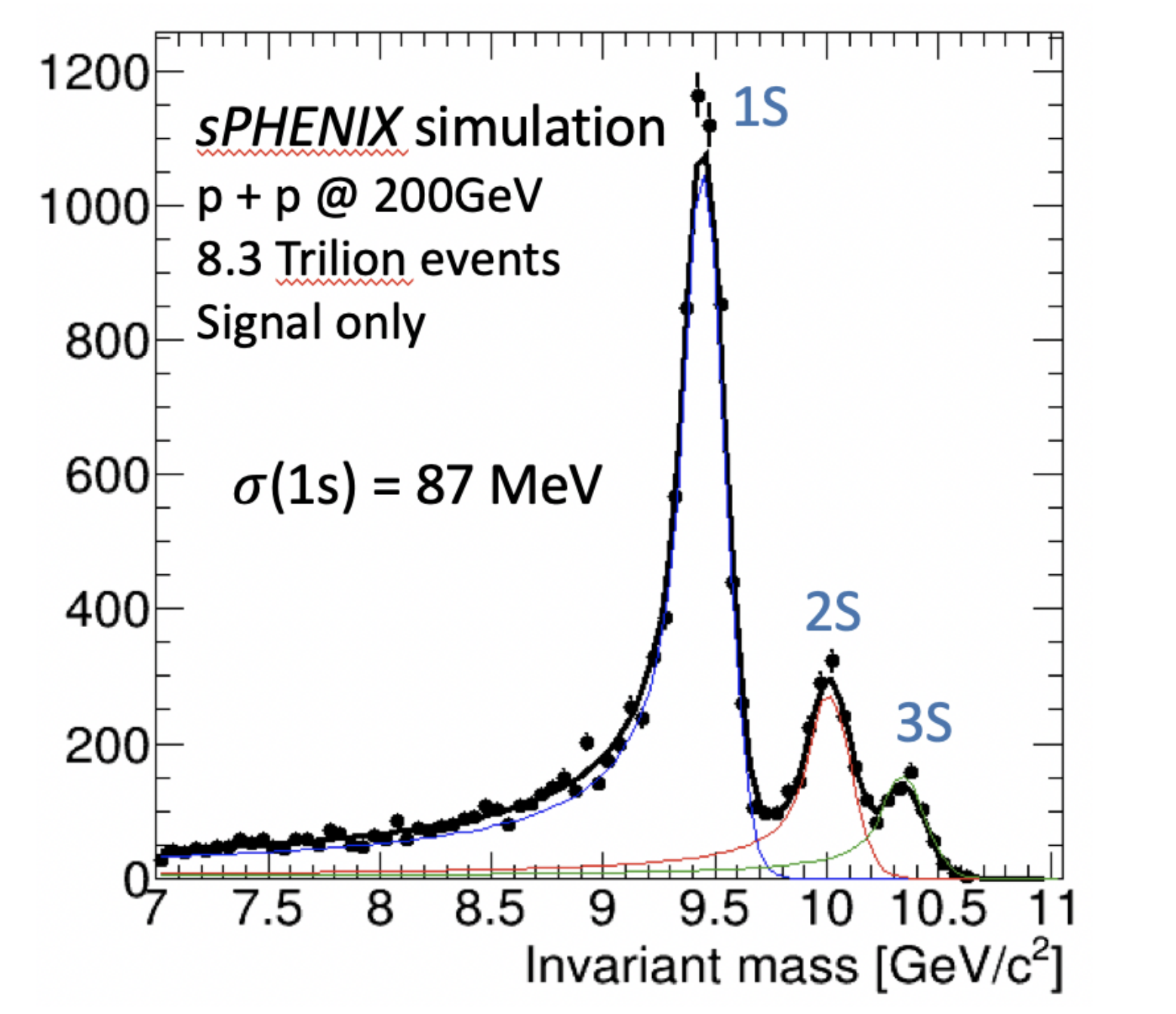}
	\caption{An example invariant mass spectrum for upsilon mesons reconstructed in sPHENIX simulations. The upsilon spectroscopy physics program largely drives the resolution requirements of the tracking.}
	\label{fig:upsilons}
\end{figure}

The physics requirements for tracking are largely driven by the goals to reconstruct the $\Upsilon(1S)$, $(2S)$, and $(3S)$ states, measuring large transverse momentum jets, and jet substructure. To measure the three upsilon states, $e^+e^-$ pairs from upsilon decays must be resolved with a mass resolution of less than 100 MeV. Therefore, tracks with a momentum of 4-8 GeV must have a resolution of approximately 1.2\%. To resolve high momentum tracks for jet substructure measurements, the tracking must have a resolution at high $p_T$ of approximately $\Delta p/p\simeq 0.2\%\cdot p$. Figure~\ref{fig:upsilons} shows an example mass spectrum of the three upsilon states as reconstructed in $\sqrt{s}=200$ GeV $p$$+$$p$ collisions. In addition to these requirements, the tracking must be robust against large combinatoric background environments, particularly within the TPC. Since the drift time is longer than the nominal bunch spacing provided by RHIC, there is potential for significant out-of-time pileup sampled in the TPC. 

\begin{figure}[!thb]
	\centering
	\includegraphics[width=0.494\textwidth]{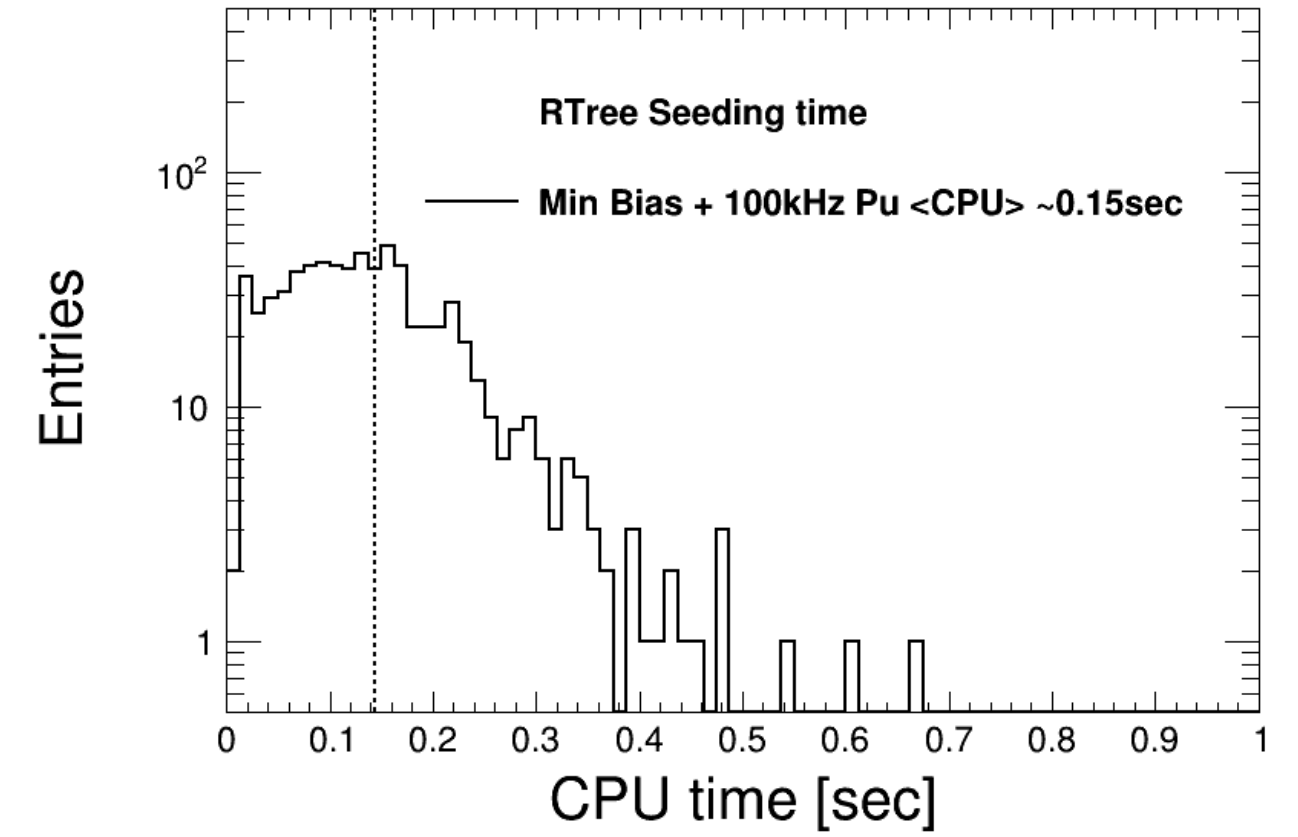}
	\includegraphics[width=0.494\textwidth]{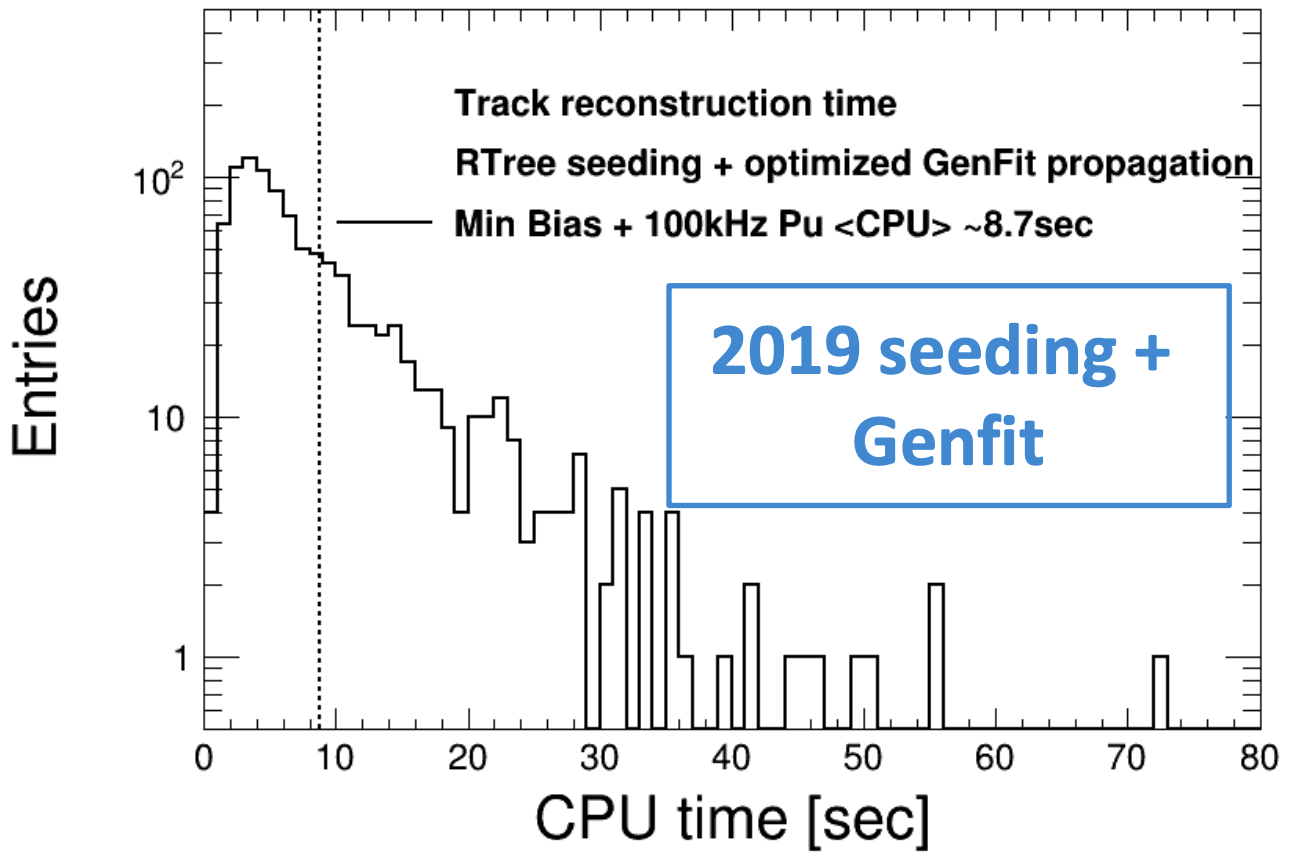}
	\caption{The CPU time for RTree seeding (left). The total CPU time for the track fitting procedure in the current sPHENIX tracking framework (right). Note the different scales of the x axes in the figures. }
	\label{fig:seeds}
\end{figure}

Currently, sPHENIX uses Hough track seeding and the GenFit2~\cite{genfit} package to perform track propagation and track fitting. In particular, the Hough seeding takes $\mathcal{O}(100)$ seconds per event, which is too slow for sPHENIX's computing needs. The collaboration is actively exploring Cellular Automaton seeding using RTrees, which is a geometric indexing that provides seeds based on nearest neighbors. The left panel in Fig.~\ref{fig:seeds} shows the seeding CPU time per minimum bias plus 100 kHz pileup event when using RTrees only; this improves the seeding time by several orders of magnitude compared to the Hough seeding. Current efforts to implement Cellular Automaton indicate further improvements. The right panel of Fig.~\ref{fig:seeds}
shows the total time to perform track fitting in GenFit with the RTree seeding implementation. The current tracking algorithm gives approximately 9 seconds per event to perform the full track fit. Therefore, to meet the desired goal of 5~seconds per event given the challenging experimental conditions in which sPHENIX will operate, sPHENIX is exploring the ACTS software package.

\section{ACTS Implementation in sPHENIX}

ACTS is a software package that is being developed by members of several different particle physics collaborations. ACTS is intended to be an experiment-independent set of track reconstruction tools written in modern \texttt{C++} that is performant yet customizable. The toolkit development was largely motivated by the High-Luminosity Large Hadron Collider (HL-LHC) that will begin data taking in 2027. At the HL-LHC, ATLAS and CMS expect environments in which roughly 200 $p$$+$$p$ collisions per bunch crossing occur~\cite{hllhc}. Therefore, the speed at which data are processed must be dramatically improved to accommodate the significantly larger data rates. This, in addition to modernizing certain tracking code, prompted the development of ACTS. sPHENIX expects roughly comparable hit occupancies with three to eight heavy ion collisions per bunch crossing as what is expected at the HL-LHC; thus, ACTS is a natural candidate for tracking in the types of environments expected at sPHENIX.

\begin{figure}[!tbh]
	\centering
	\includegraphics[width=0.4\textwidth]{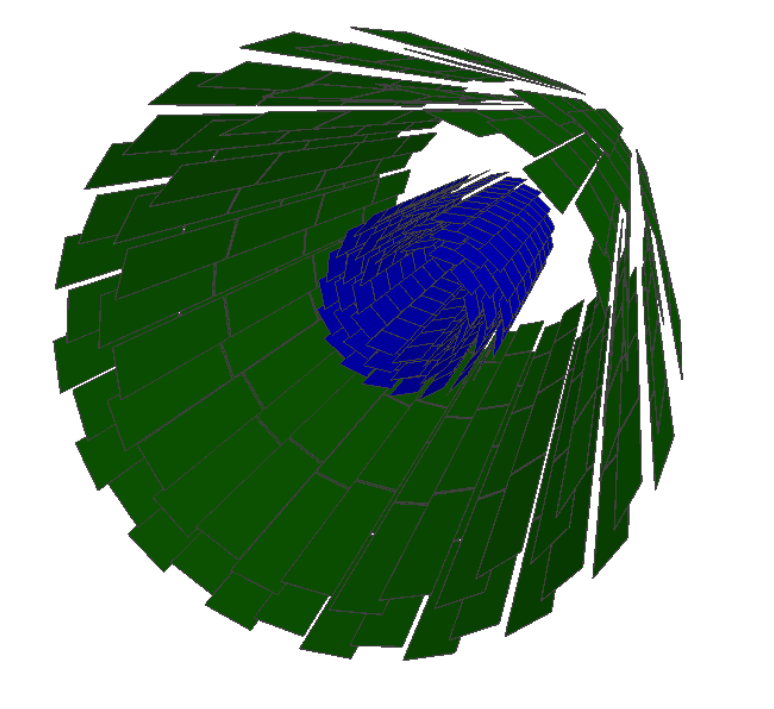}
	\caption{The sPHENIX MVTX and INTT silicon detectors as implemented within ACTS.}
	\label{fig:actssilicon}
\end{figure}

The first step for implementing ACTS into the sPHENIX software stack is to properly translate the tracking detector geometry into the expected ACTS geometry. The main detector element used for track fitting is the \texttt{Acts::Surface}. ACTS has an available ROOT \texttt{TGeometry} plugin that can take the relevant active \texttt{TGeo} objects and convert them into \texttt{Acts::Surfaces}. Since sPHENIX already has a detailed and well-tested {\sc{GEANT4}} geometry description that uses the ROOT \texttt{TGeoManager}, this plugin was a natural choice. Figure~\ref{fig:actssilicon} shows the MVTX and INTT active silicon surfaces as implemented within ACTS. The geometry is imported directly from the \texttt{TGeoManager}, so any changes in the sPHENIX {\sc{GEANT4}} description will be automatically propagated to the \texttt{Acts::Surface} description.

\begin{figure}[!tbh]
	\centering
	\includegraphics[width=0.5\textwidth]{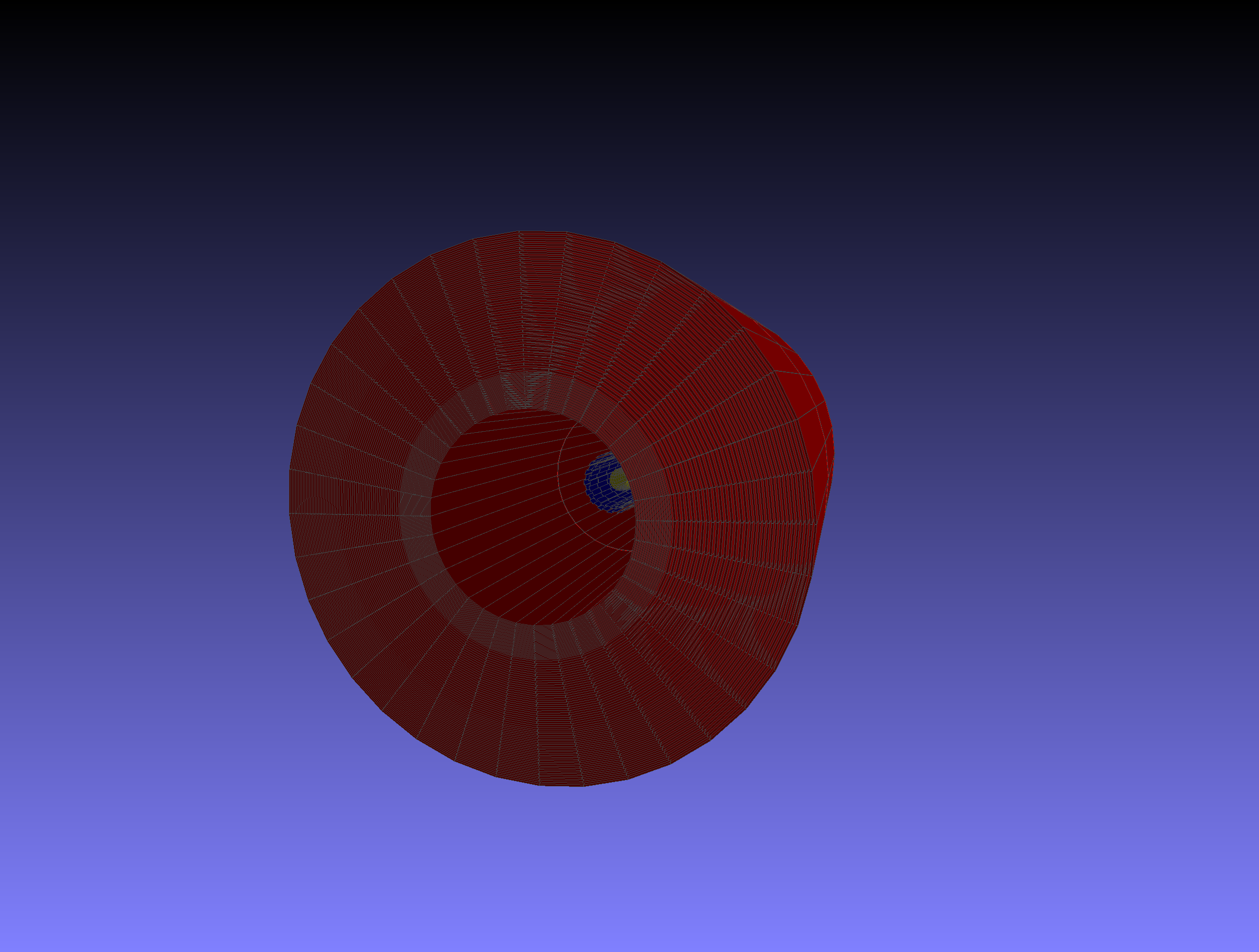}
	\caption{The sPHENIX TPC as currently implemented within ACTS. The TPC surfaces are shown in red, and the MVTX and INTT surfaces are shown in yellow and blue, respectively.}
	\label{fig:tpc}
\end{figure}

The TPC implementation is not as straightforward because, at the time of writing, ACTS does not support continuous volume geometries for tracking. This is because track fitting is performed measurement-by-measurement on the local surfaces; this is not possible for a TPC since clusters can be formed anywhere within the continuous TPC volume. The implementation of global track fitting is an ongoing effort within ACTS to allow the use of drift chamber and TPC geometries. In the meantime, the sPHENIX TPC is modeled as a set of layered surfaces that correspond to the TPC readout layers. This models the TPC as a set of concentric cylindrical layers in which each cylinder is divided into surfaces that span half the TPC length in $z$ and 10$^\circ$ in azimuthal angle. Figure~\ref{fig:tpc} shows this current TPC implementation within ACTS.

To provide ACTS with the information necessary for performing the track fitting, the sPHENIX software maps the relevant sPHENIX tracking objects to the analogous ACTS tracking objects. These maps are maintained so that the sPHENIX objects can be passed to ACTS, and then the returned ACTS objects can be mapped back to their corresponding sPHENIX counterparts. For example, the primary class responsible for cluster or hit information is the \texttt{Acts::SourceLink}. Therefore, a map is created that links a sPHENIX cluster, or \texttt{TrkrCluster}, to the corresponding \texttt{Acts::SourceLink} with the same hit and surface information. The software is written in a modular way so that each step of data preparation for ACTS is contained within its own sPHENIX module. This allows for flexibility in the track fitting process so that any of the various seeding, propagating, or fitting algorithms can be swapped in and out to test the functionality of various combinations.

\section{Conclusions}

The sPHENIX experiment is a dedicated jet and heavy-flavor experiment that is being constructed at RHIC. sPHENIX will measure a variety of strong force interaction physics in proton-proton, proton-nucleus, and nucleus-nucleus collisions. To achieve the physics goals, precise tracking must be implemented that can withstand large backgrounds from out-of-time pileup collisions in all three collision systems. Additionally, the data must be reconstructed in a timely manner on a fixed computational center available at BNL. To reach the current tracking budget of 5 seconds per event in minimum bias heavy ion collisions with 100 kHz pileup, sPHENIX is implementing the ACTS tracking reconstruction toolkit. The sPHENIX geometry has been appropriately implemented, and the the track fitter performance is being studied. Future results will identify whether ACTS will meet the computational goals and physics requirements for sPHENIX.


\Acknowledgements
This work has been supported by the Office of Nuclear Physics in the US Department of Energy (DOE) Office of Science. Oak Ridge National Laboratory is managed by UT-Battelle LLC for DOE under contract no. DE-AC05-00OR22725.




\begin{thebibliography}{99}


\bibitem{sphenix}
	sPHENIX Collaboration, ``sPHENIX: An Upgrade Proposal,'' \href{https://arxiv.org/abs/1501.06197}{arXiv:1501.06197}, (2015).

\bibitem{LRP}
		DOE Office of Science, 
		``Reaching for the Horizon: The 2015 Long Range Plan for Nuclear Science,''
		(2015).
		
\bibitem{sphenixcoldqcd}
	sPHENIX Collaboration. ``Medium-energy nuclear physics measurements with the sPHENIX Barrel,'' \href{https://indico.bnl.gov/event/3866/}{https://indico.bnl.gov/event/3866/} (2017).
\bibitem{acts}
	A Common Tracking Software (ACTS) package, release-0.23.X,
	\href{http://github.com/acts-project/acts}{http://github.com/acts-project/acts}.

\bibitem{actstext}
	Xiaocong Ai (for the ACTS Developers), ``Acts: A common tracking software,'' \href{https://arxiv.org/abs/1910.03128}{arXiv:1910.03128} (2019).

\bibitem{sphenixtdr}
	sPHENIX Collaboration. ``sPHENIX Technical Design Report,'' \href{https://indico.bnl.gov/event/7081/}{https://indico.bnl.gov/event/7081/} (2019).

\bibitem{genfit}
	GenFit Tracking package, \href{http://github.com/GenFit/GenFit}{http://github.com/GenFit/GenFit}.
	
\bibitem{hllhc}
	G. Apollinari \textit{et al.}, ``High-Luminosity Large Hadron Collider (HL-LHC): Preliminary Design Report,'' \href{https://cds.cern.ch/record/2116337}{https://cds.cern.ch/record/2116337}.
	
	
\end{thebibliography}
\end{document}